\documentclass[10pt]{article}
\usepackage{graphicx}

\def\Title#1{\begin{center} {\Large #1 } \end{center}}
\def\Author#1{\begin{center}{ \sc #1} \end{center}}
\def\Address#1{\begin{center}{ \it #1} \end{center}}

\newcommand\pubblock{\rightline{\begin{tabular}{l} Proceedings of the Second Annual LHCP\\ \pubnumber\\
         \pubdate  \end{tabular}}}

\newenvironment{Abstract}{\begin{quotation} \begin{center} 
             \large ABSTRACT \end{center}\bigskip 
      \begin{center}\begin{large}}{\end{large}\end{center} \end{quotation}}

\newenvironment{Presented}{\begin{quotation} \begin{center} 
             PRESENTED AT\end{center}\bigskip 
      \begin{center}\begin{large}}{\end{large}\end{center} \end{quotation}}

\def\beq{\begin{equation}}
\def\eeq#1{\label{#1}\end{equation}}
\def\eeqn{\end{equation}}

\def\beqa{\begin{eqnarray}}
\def\eeqa#1{\label{#1}\end{eqnarray}}
\def\eeqan{\end{eqnarray}}

\let\bar=\overbar

\def\L{{\cal L}}

\def\Dslash{\not{\hbox{\kern-4pt $D$}}}
\def\dslash{\not{\hbox{\kern-2pt $\del$}}}

\def\msb{{\bar{\ssstyle M \kern -1pt S}}}

\usepackage{color}

\newcommand\pubnumber{  }

\newcommand\pubdate{\today}

\def\affiliation{
University of Rzesz\'ow,\\
PL-35-959 Rzesz\'ow, Poland}

\def\support{\footnote{
Work supported by the Polish
NCN grant DEC-2013/09/D/ST2/03724
}}

\begin{document}

\large
\begin{titlepage}
\pubblock

\vfill
\Title{  Production of $W^+W^-$ pairs via subleading processes at the LHC  }
\vfill

\Author{ Marta Luszczak \support }
\Address{\affiliation}
\vfill
\begin{Abstract}

We discuss many new subleading processes for inclusive production of 
$W^+ W^-$ pairs generally not included in the literature so far.
We focus mainly on photon-photon induced processes.
We include elastic-elastic, elastic-inelastic, inelastic-elastic 
and inelastic-inelastic contributions. 
Predictions for the total cross section and differential
distributions in $W$- boson rapidity and transverse momentum as well
as $WW$ invariant mass are presented. The $\gamma \gamma$ components
only constitute about 1-2 \% of the inclusive $W^+ W^-$ cross section
but increases up to about 10 \% at large $W^{\pm}$ transverse momenta, and are even
comparable to the dominant $q \bar q$ component at large $M_{WW}$,
i.e. they are much larger than the $g g \to W^+ W^-$ one.

\end{Abstract}
\vfill

\begin{Presented}
The Second Annual Conference\\
 on Large Hadron Collider Physics \\
Columbia University, New York, U.S.A \\ 
June 2-7, 2014
\end{Presented}
\vfill
\end{titlepage}
\def\thefootnote{\fnsymbol{footnote}}
\setcounter{footnote}{0}
%

\normalsize 


\section{Introduction}

The $\gamma \gamma \to W^+ W^-$ process is interesting 
by itself as it can be used to test the Standard Model and any 
other theories beyond the Standard Model.
The photon-photon contribution for the purely exclusive production 
of $W^+ W^-$ was considered recently in the literature 
\cite{royon,piotrzkowski}.
The exclusive diffractive mechanism of central exclusive production
of $W^+W^-$ pairs in proton-proton collisions at the LHC 
(in which diagrams with an intermediate virtual Higgs boson as well as quark box
diagrams are included) was discussed in Ref.~\cite{LS2012} and turned 
out to be negligibly small.
The diffractive production and decay of the Higgs boson into the $W^+W^-$ pair 
was also discussed in Ref.~\cite{WWKhoze}. 
The $W^+W^-$ pair production signal would
be particularly sensitive to New Physics contributions in 
the $\gamma \gamma \to W^+ W^-$ subprocess \cite{royon,piotrzkowski}. 
Corresponding measurements would be possible to be performed at ATLAS or CMS
provided the very forward proton detectors are installed
\cite{forward_protons}. 
This would be a valuable supplement of the so far accepted scientific
program.
We concentrate on inclusive production 
of $W^+ W^-$ pairs.
The inclusive production of $W^+ W^-$ has been measured recently with 
the CMS and ATLAS detectors \cite{CMS2011, ATLAS2012}.
The total measured cross section with the help of the CMS detector is  
41.1 $\pm$ 15.3 (stat) $\pm$ 5.8 (syst) $\pm$ 4.5 (lumi) pb, 
the total measured cross section with the ATLAS detector with slightly
better statistics is
54.4 $\pm$  4.0 (stat.) $\pm$  3.9 (syst.) $\pm$  2.0 (lumi.) pb. 
The more precise ATLAS result is somewhat bigger than the Standard Model
predictions of 44.4  $\pm$  2.8 pb \cite{ATLAS2012}.
The Standard Model predictions do not include several potentially
important subleading processes.
We  review several  processes which
have been ignored in the present Standard Model predictions.

\section{Inclusive production of $W^+W^-$ pairs}
\label{sec:inclusive}

The dominant contribution of $W^+W^-$ pair production is initiated by
quark-antiquark annihilation \cite{DDS95}.
The gluon-gluon contribution to the inclusive cross section 
was calculated first in Ref.~\cite{gg_WW}.

Therefore in the following for a comparison we also consider
quark-antiquark and gluon-gluon components to the inclusive 
cross section. They will constitute a reference point for 
our calculations of the two-photon contributions. 

\subsection{$\gamma \gamma \to W^+ W^-$ mechanism}

In this section, we discuss the inclusive $\gamma \gamma \to W^+ W^-$
induced mechanisms. 
We shall calculate the contribution to the inclusive
$p p \to W^+ W^- X$ process for the first time in the literature.

\begin{figure*}
\begin{center}
\includegraphics[width=3.5cm]{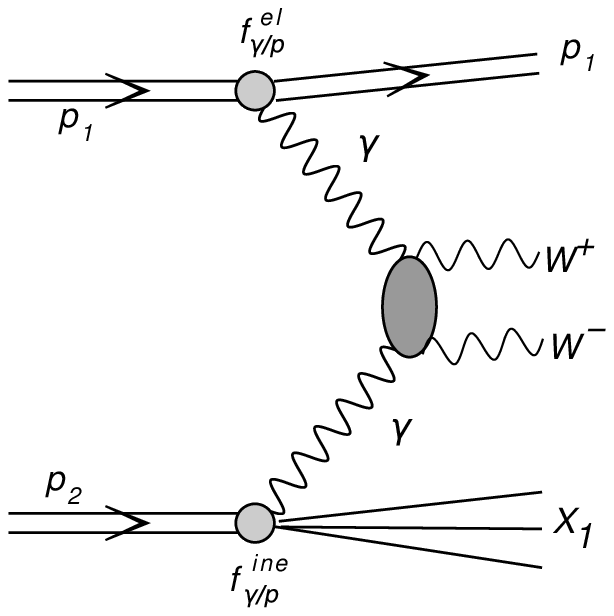}
\includegraphics[width=3.5cm]{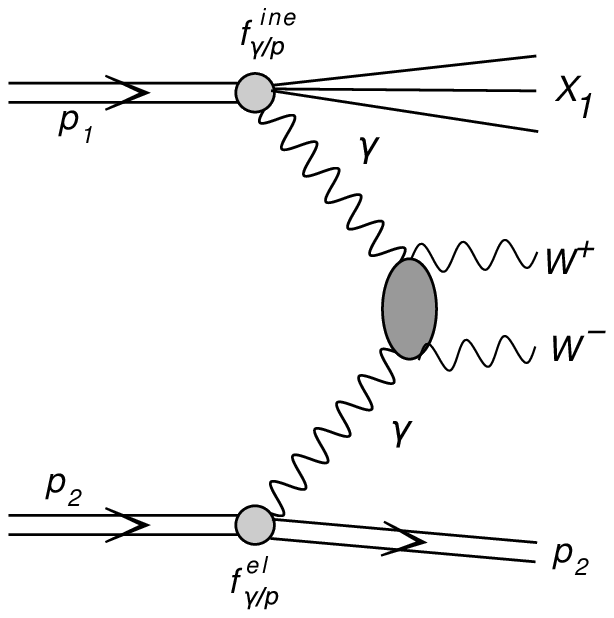}
\includegraphics[width=3.5cm]{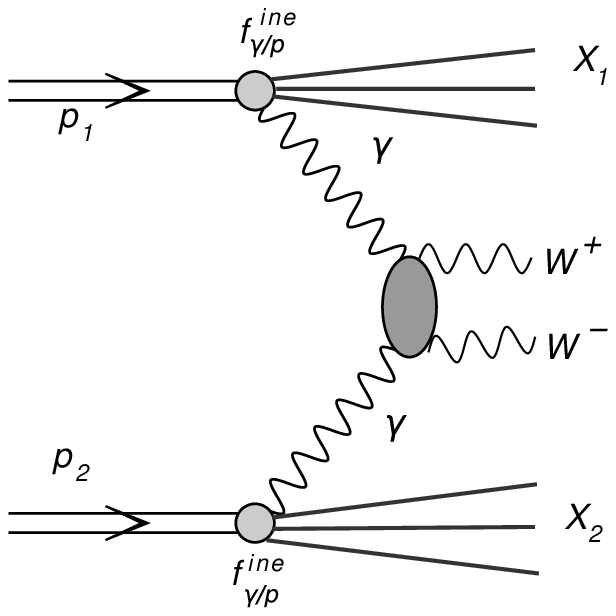}
\caption{Diagrams representing inelastic photon-photon induced mechanisms 
for the production of $W^+ W^-$ pairs.
}
\label{fig:new_diagrams}
\end{center}
\end{figure*}

If at least one photon is a constituent of the nucleon then 
the mechanisms presented in Fig.\ref{fig:new_diagrams} are possible.
In these cases at least one of the participating protons does not survive 
the $W^+ W^-$ production process. In the following we consider 
two different approaches to the problem.
An approach how to include photons into inelastic processes 
was proposed some time ago by Martin, Roberts, 
Stirling and Thorne in Ref. \cite{MRST04}. In their approach the photon 
is treated on the same footing as quarks, antiquarks and gluons.
They proposed a QED-corrected evolution equation for the parton 
distributions of the proton \cite{MRST04}.

In leading order approximation the corresponding triple differential cross section 
for the inelastic-inelastic photon-photon contribution can be written as
usually in the parton-model formalism:
\begin{eqnarray}
\frac{d \sigma^{\gamma_{in} \gamma_{in}}}{d y_1 d y_2 d^2p_t} &=& \frac{1}{16 \pi^2 {\hat s}^2}
x_1 \gamma_{in}(x_1,\mu^2) \; x_2 \gamma_{in}(x_2,\mu^2) \;
\overline{|{\cal M}_{\gamma \gamma \to W^+W^-}|^2} \; .
\end{eqnarray}

The above contribution includes only cases when both nucleons do not survive
the collision and the nucleon debris is produced instead. The case when
at least one nucleon survives the collision has to be considered
separately. Corresponding contributions to the cross section 
can then be written as:
\begin{eqnarray}
\frac{d \sigma^{\gamma_{in} \gamma_{el}}}{d y_1 d y_2 d^2p_t} &=& \frac{1}{16 \pi^2 {\hat s}^2}
x_1 \gamma_{in}(x_1,\mu^2) \; x_2 \gamma_{el}(x_2,\mu^2) \;
\overline{|{\cal M}_{\gamma \gamma \to W^+W^-}|^2} \; ,\nonumber \\
\frac{d \sigma^{\gamma_{el} \gamma_{in}}}{d y_1 d y_2 d^2p_t} &=& \frac{1}{16 \pi^2 {\hat s}^2}
x_1 \gamma_{el}(x_1,\mu^2) \; x_2 \gamma_{in}(x_2,\mu^2) \;
\overline{|{\cal M}_{\gamma \gamma \to W^+W^-}|^2} \; ,\nonumber \\
\frac{d \sigma^{\gamma_{el} \gamma_{el}}}{d y_1 d y_2 d^2p_t} &=& \frac{1}{16 \pi^2 {\hat s}^2}
x_1 \gamma_{el}(x_1,\mu^2) \; x_2 \gamma_{el}(x_2,\mu^2) \; 
\overline{|{\cal M}_{\gamma \gamma \to W^+W^-}|^2} \; , \\ 
\label{subleading_contributions}
\end{eqnarray}
for the inelastic-elastic, elastic-inelastic and elastic-elastic
components, respectively.
In the following the elastic photon fluxes are calculated using 
the Drees-Zeppenfeld parametrization \cite{DZ}, where a simple 
parametrization of the nucleon electromagnetic form factors is used.

In the case of resolved photons, the ``photonic'' quark/antiquark 
distributions in a proton must be calculated first. This can be done 
by the convolution
\begin{equation}
f_{q/p}^{\gamma} = f_{\gamma/p} \otimes f_{q/\gamma}
\end{equation}
which mathematically means:
\begin{equation}
x f_{q/p}^{\gamma}(x) = \int_x^1 d x_{\gamma} f_{\gamma/p}(x_{\gamma},\mu_s^2) 
\left( \frac{x}{x_{\gamma}} \right) 
f \left( \frac{x}{x_{\gamma}}, \mu_h^2 \right)  \; . 
\label{convolution_resolved}       
\end{equation}

Diffractive processes for $W^+ W^-$ production were not considered so
far in the literature but are potentially very important.

\begin{figure*}
\begin{center}
\includegraphics[width=3cm]{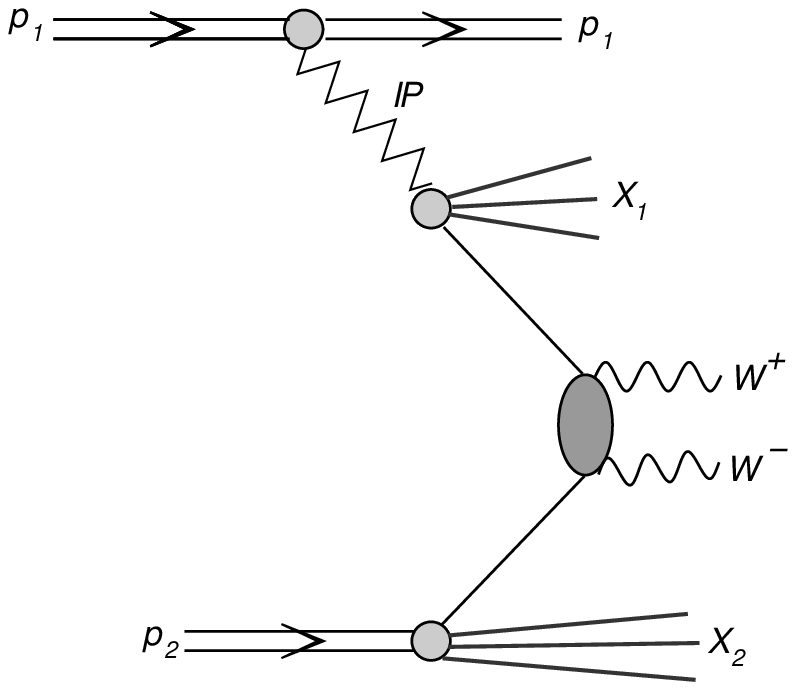}
\includegraphics[width=3cm]{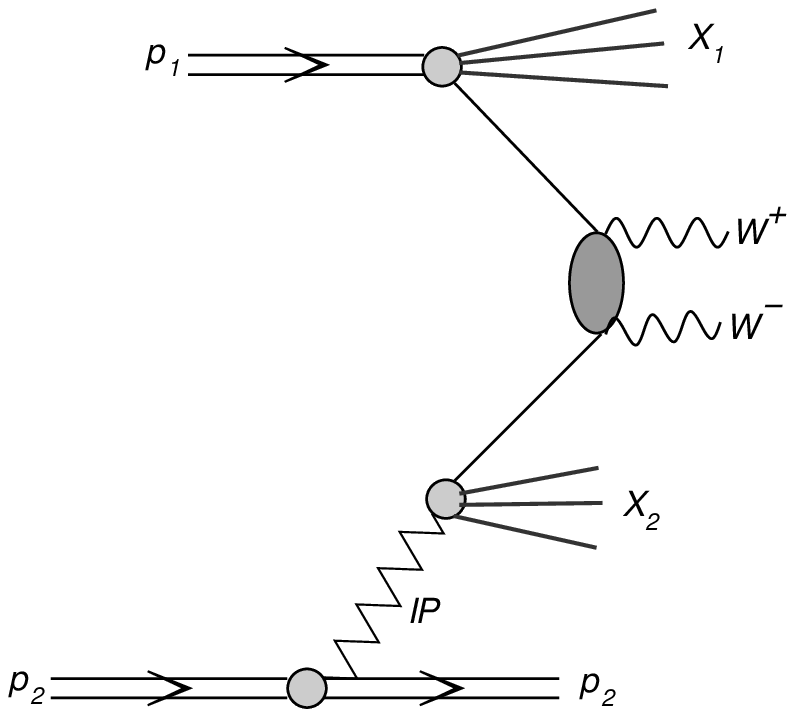}
\caption{Diagrams representing single diffractive mechanism
of the production of $W^+ W^-$ pairs.
}
\label{fig:single_diffractive}
\end{center}
\end{figure*}

In this approach one assumes that the Pomeron has a
well defined partonic structure, and that the hard process
takes place in a Pomeron--proton or proton--Pomeron (single diffraction) 
or Pomeron--Pomeron (central diffraction) processes.
The mechanism of single diffractive production of $W^+ W^-$ pairs
is shown in Fig.\ref{fig:single_diffractive}.

In the present analysis we consider both pomeron and subleading reggeon
contributions. The corresponding diffractive quark distributions
are obtained by replacing the pomeron flux by the reggeon flux and
quark/antiquark distributions in the pomeron by their counterparts
in subleading reggeon(s). The other details can be found in \cite{H1}.
In the case of pomeron exchange the upper limit in the integration
over the momentum fraction carried by the pomeron/reggeon
in the convolution formula is 0.1 for pomeron and 0.2 
for reggeon exchange.

Up to now we have assumed Regge factorization which is known
to be violated in hadron-hadron collisions.
It is known that these are soft interactions which lead to an extra 
production of particles which fill in the rapidity gaps related 
to pomeron exchange.

If rapidity gap (gaps) is required (measured) then one has to include
absorption effects in the formalism of the resolved pomeron/reggeon
which can be interpreted as a probability of no extra soft interactions
leading to a distruction of rapidity gap.

\begin{figure*}
\begin{center}
\includegraphics[width=3cm]{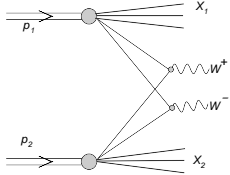}
\caption{Diagram representing double parton scattering mechanism
of the production of $W^+ W^-$ pairs.
}
\label{fig:DPS}
\end{center}
\end{figure*}

The diagram representating the double parton scattering
process is shown in Fig.\ref{fig:DPS}.

The cross section for double parton scattering is often modelled
in the factorized anzatz which in our case would mean:
\begin{equation}
\sigma_{W^+ W^-}^{DPS} = \frac{1}{\sigma_{qq}^{eff}} 
\sigma_{W^{+}}
\sigma_{W^{-}}
\; .
\label{factorized_model}
\end{equation}

The factorized model (\ref{factorized_model}) can be generalized 
to more differential distributions (see e.g. \cite{LMS2012,MS2013}).
For example in our case of $W^{+} W^{-}$ production the cross section
differential in $W$ boson rapidities can be written as:
\begin{equation}
\frac{d \sigma_{W^+ W^-}^{DPS}}{d y_{+} d y_{-}} =
\frac{1}{\sigma_{qq}^{eff}} 
\frac{d\sigma_W^{+}}{d y_{+}}
\frac{d\sigma_W^{-}}{d y_{-}} \; .
\label{generalized_factorized_model}
\end{equation}
%
\section{Results}

\begin{figure}
\begin{center}
\includegraphics[width=5cm]{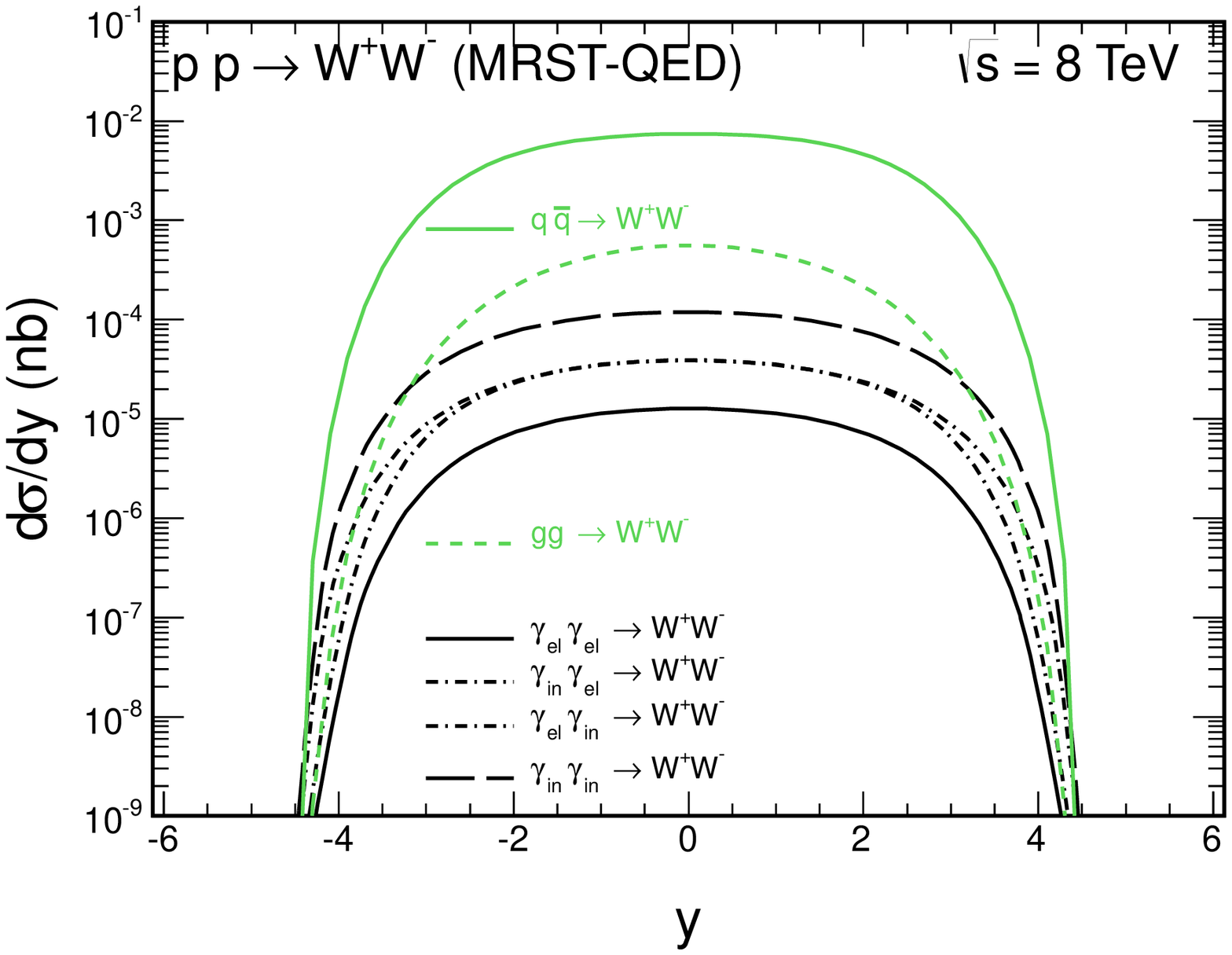}\\
\includegraphics[width=5cm]{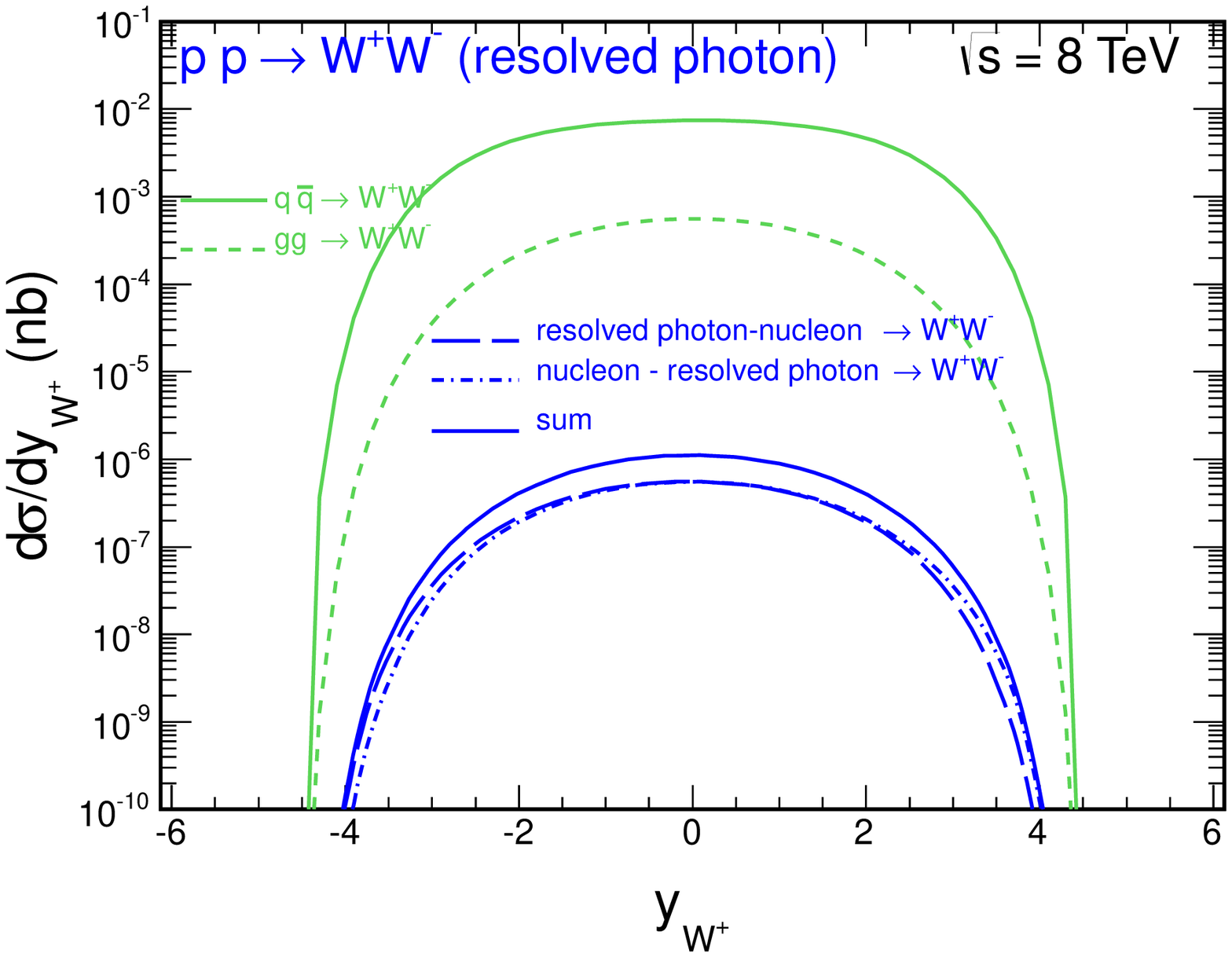}
\includegraphics[width=5cm]{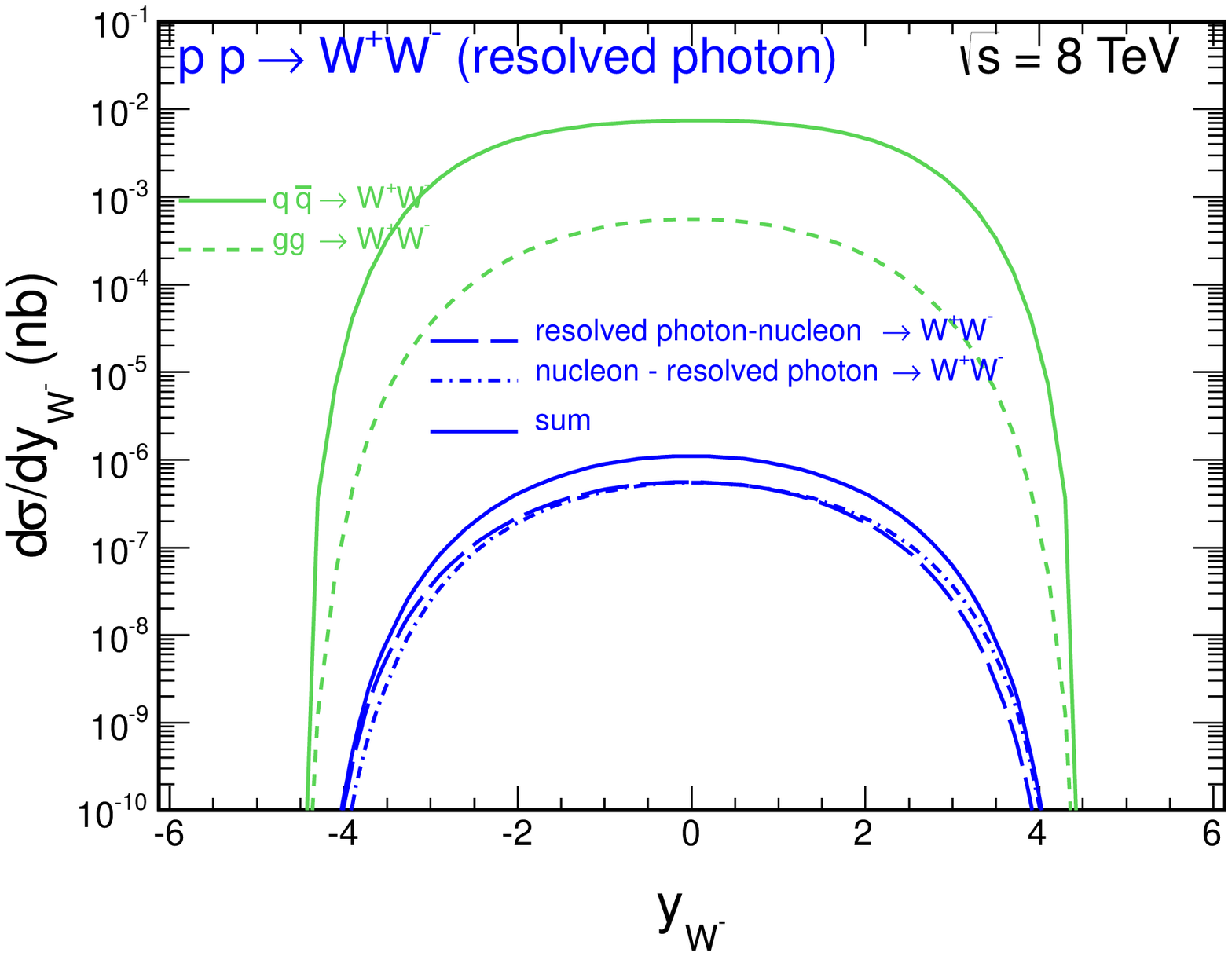}\\
\includegraphics[width=5cm]{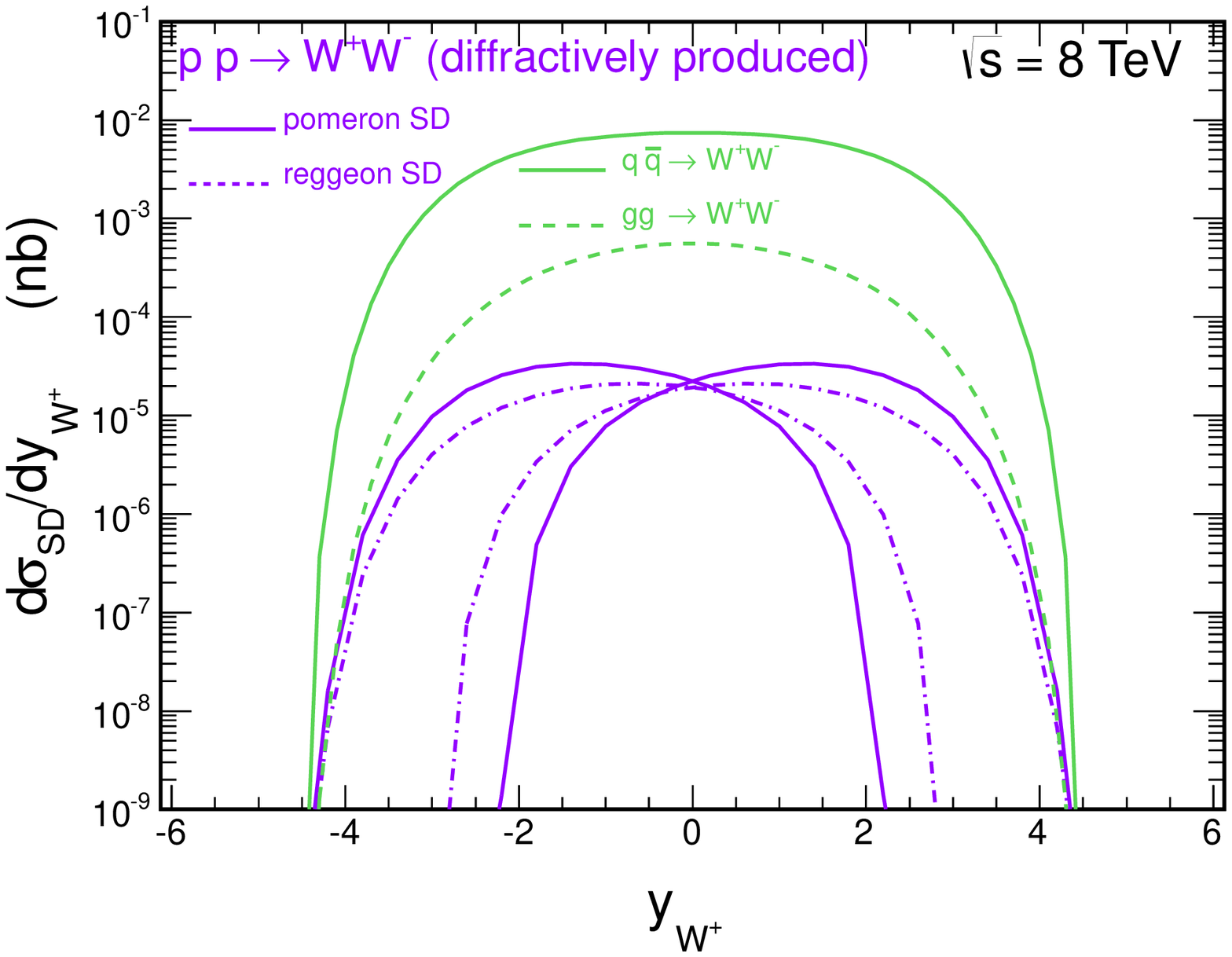}
\includegraphics[width=5cm]{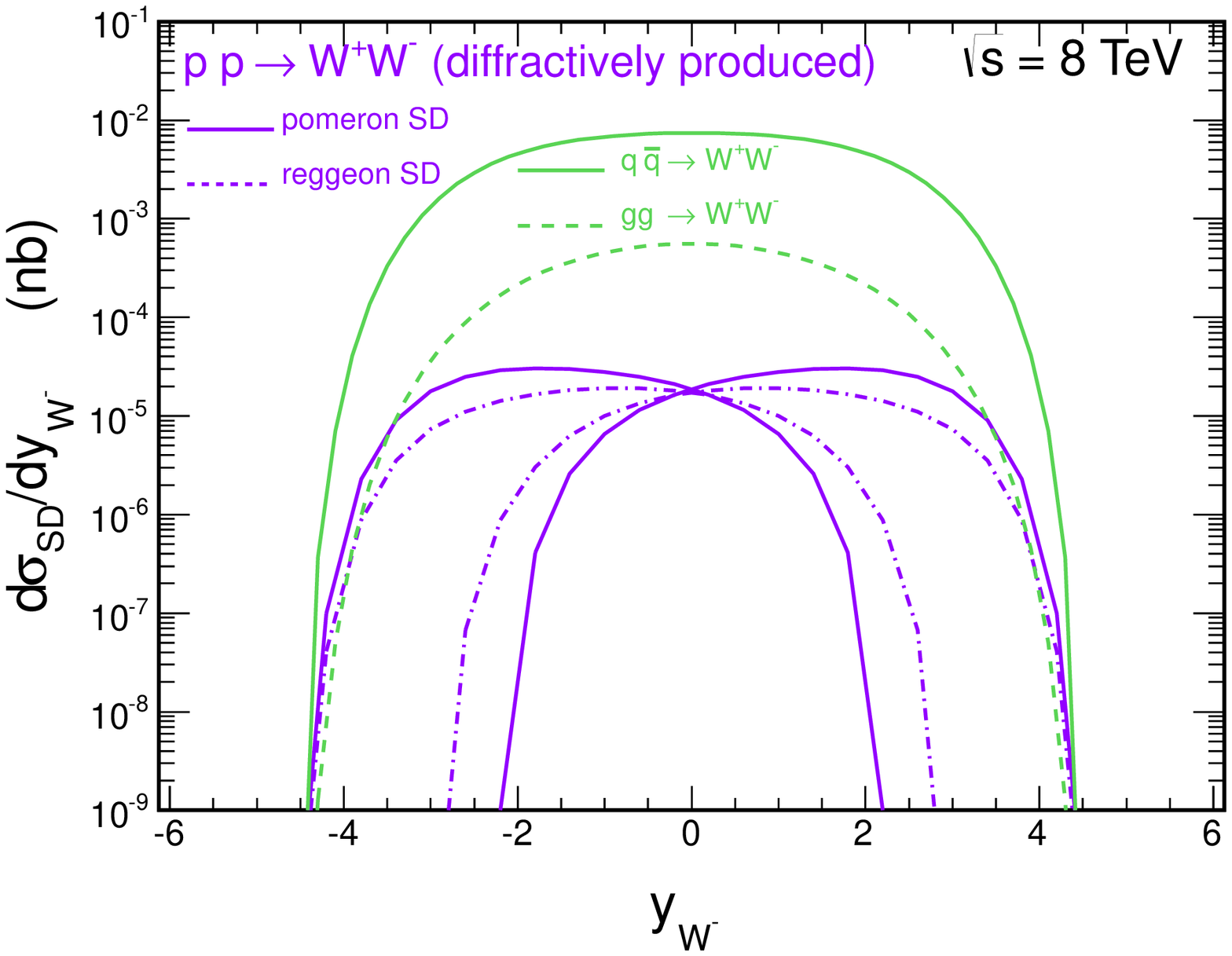}
\end{center}
\caption {$W$ boson rapidity distribution for $\sqrt{s}$ = 8 TeV.
The top panel shows contributions of all photon-photon induced processes
as a function of $y$, the
middle panels resolved photon contributions and the bottom panels 
distributions of the diffractive contribution. The diffractive
cross section has been multiplied by the gap survival factor
$S_G^2$ = 0.08 as needed for requirement of rapidity gaps.
}
\label{fig:dsig_dy}
\end{figure}

The distribution in $W$ boson rapidity is shown in
Fig.\ref{fig:dsig_dy}.
The diffractive contribution is an order of magnitude larger than
the resolved photon contribution. The estimated reggeon contribution is
of similar size as the pomeron contribution.
The distributions of $W^+$ and $W^-$ for the double-parton scattering 
contribution are different and, in 
the approximation discussed here, 
have shapes identical to those for single production of $W^+$ and $W^-$, 
respectively.
It would therefore be interesting to obtain separate distributions
for $W^+$ and $W^-$ experimentally. 

\begin{figure}
\begin{center}
\includegraphics[width=5cm]{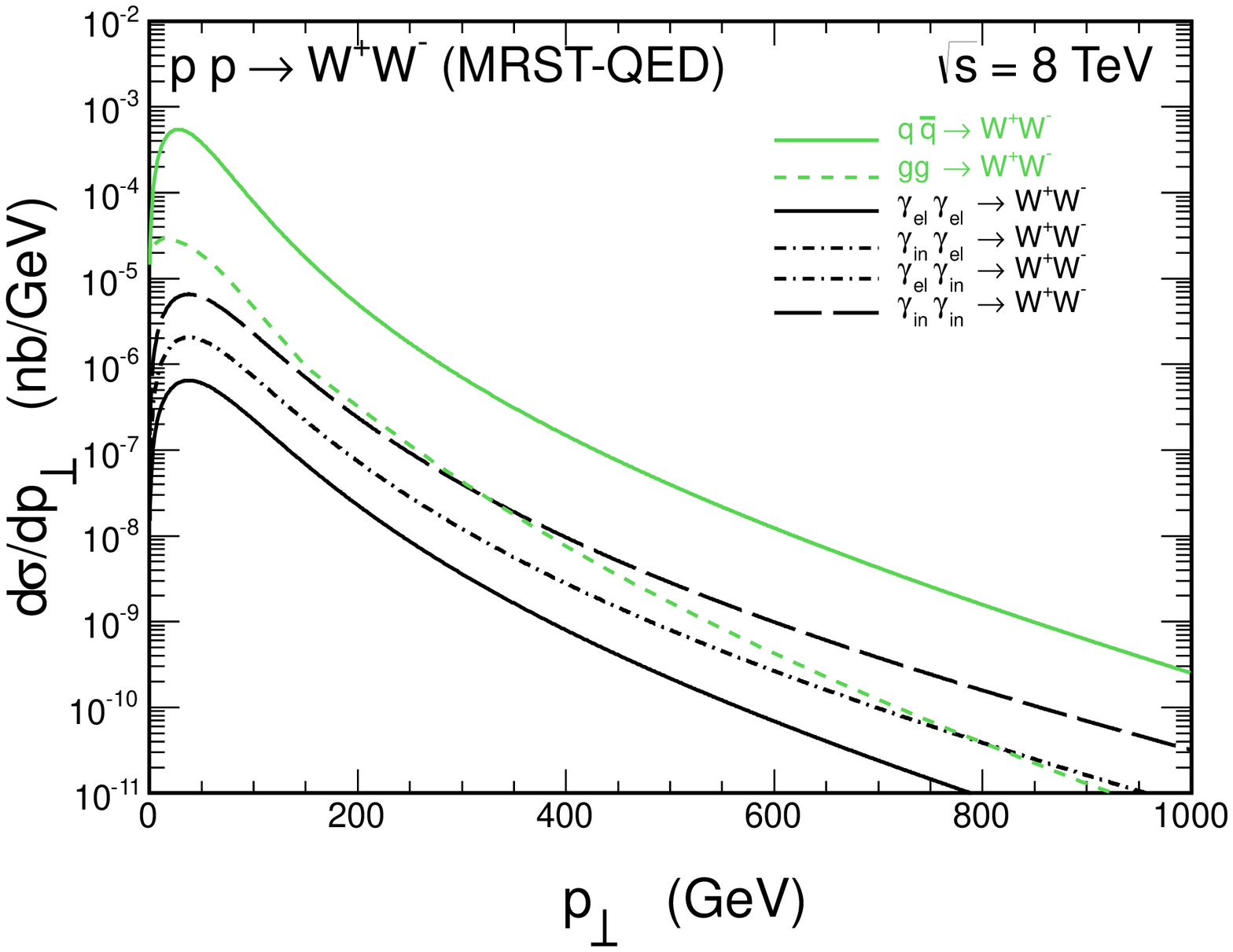}
\includegraphics[width=5cm]{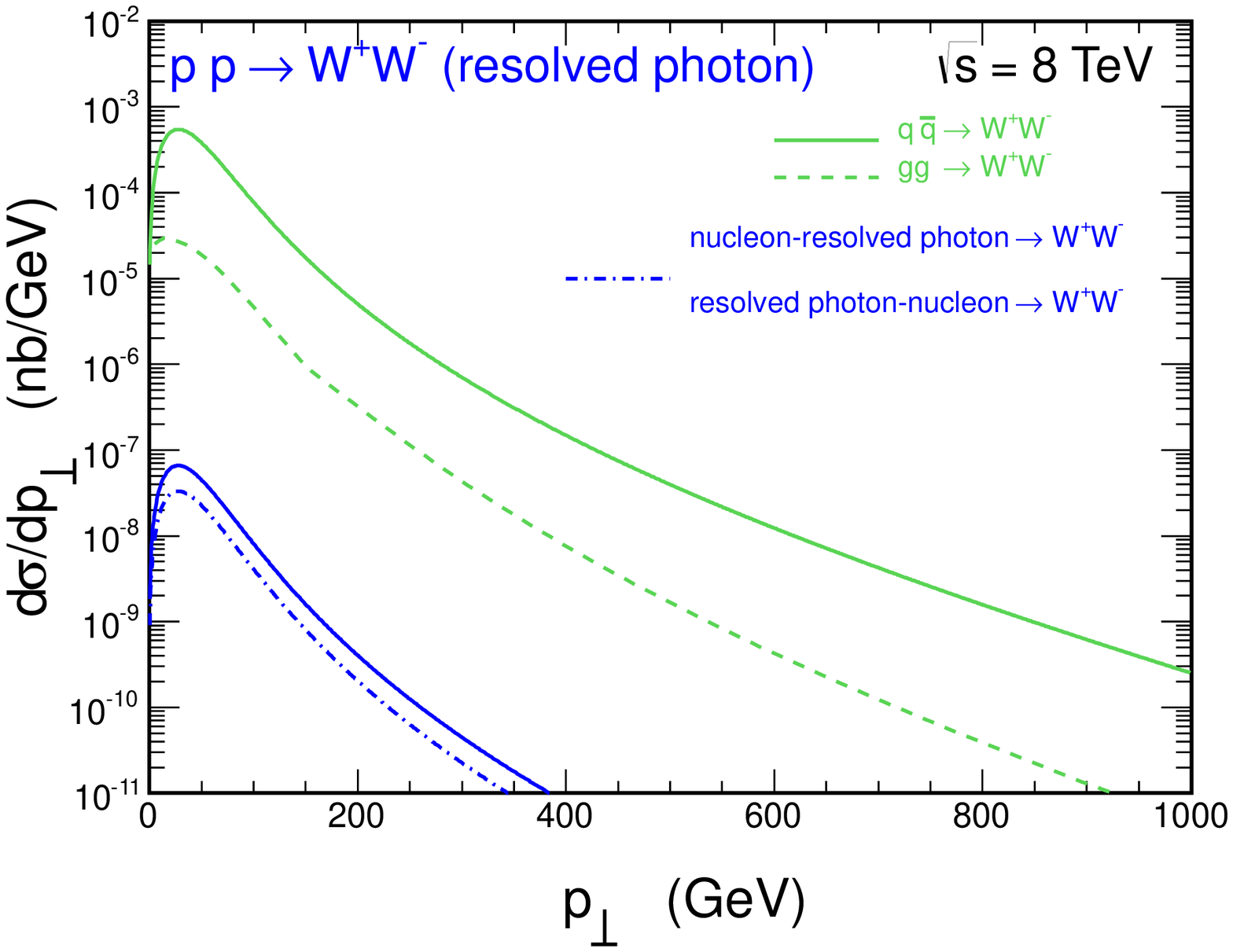}
\includegraphics[width=5cm]{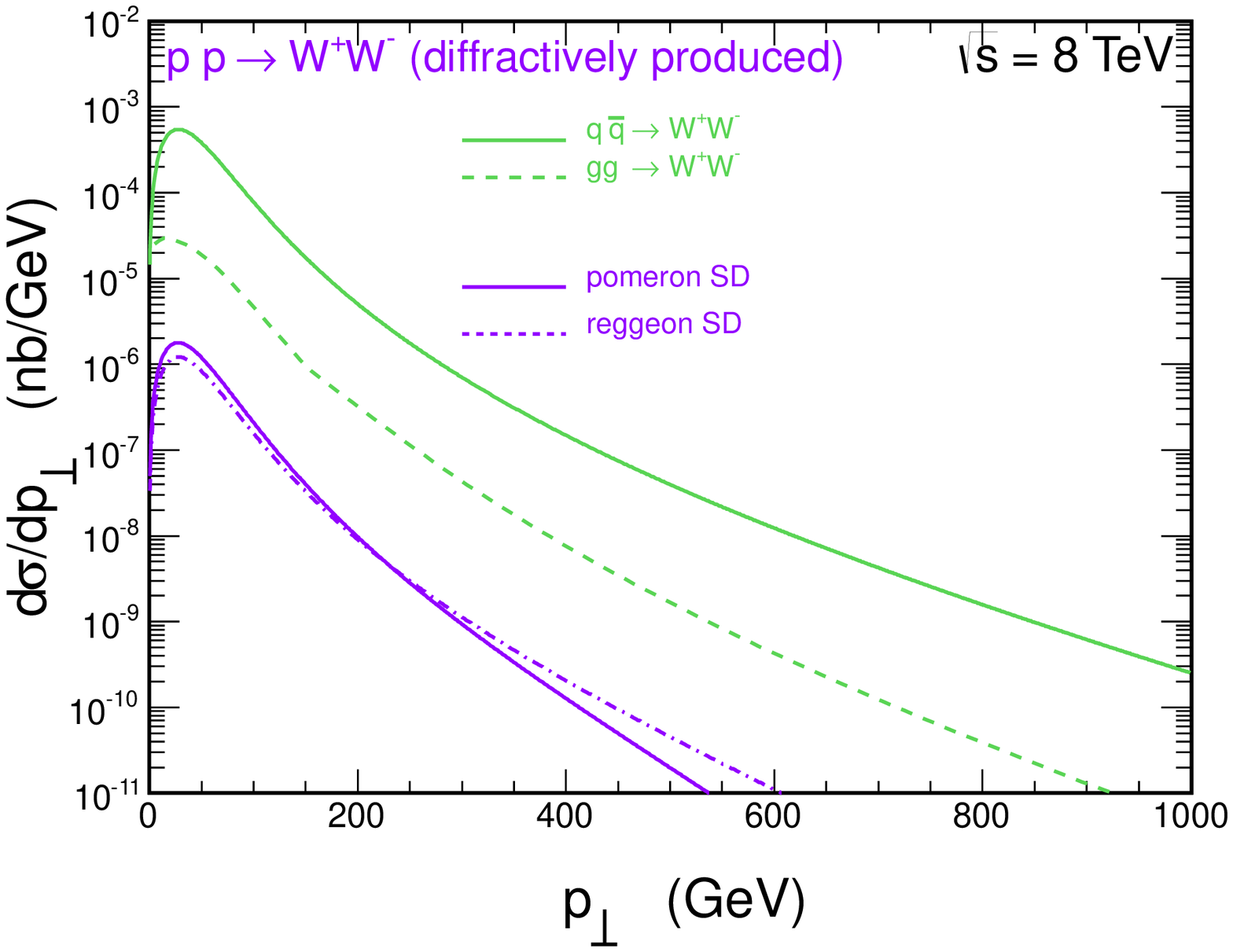}
\end{center}
\caption{ $W$ boson transverse momentum distribution for 
$\sqrt{s}$ = 8 TeV.
The left panel shows all photon-photon induced processes, the
middle panel resolved photon contributions and the right panel the
diffractive contribution.
The diffractive cross section has been multiplied by the gap survival
factor $S_G^2 =$ 0.08.
The distributions are of similar shape
except for the $\gamma \gamma$ one which is slightly harder.
}
\label{fig:dsig_dpt}
\end{figure}

In Fig.~\ref{fig:dsig_dpt} we present distributions in the transverse momentum
of $W$ bosons. All photon-photon components have rather similar shapes.
The photon-photon contributions are somewhat harder than those for
diffractive and resolved photon mechanisms. 

%

\section{Conclusions}

We have calculated for the first time a complete set of
photon-photon  and resolved photon-(anti)quark and (anti)quark-resolved 
photon contributions to the inclusive production of $W^+ W^-$ pairs. 
The photon-photon contributions can be classified into four topological
categories: elastic-elastic, elastic-inelastic, inelastic-elastic
and inelastic-inelastic, depending whether proton(s) survives (survive)
the emission of the photon or not. The elastic-inelastic and
inelastic-elastic contributions were calculated here for the first time.
The photon-photon contributions were calculated within the QCD-improved method 
using MRST(QED) parton distributions. 
The second approach was already applied to the production of
Standard Model charged lepton pair production
and $c \bar c$ production.
In the first approach we have obtained: 
$\sigma_{ela,ela} > \sigma_{ela,ine} = \sigma_{ine,ela} >
\sigma_{ine,ine}$.
In the more refined second approach we have got
$\sigma_{ela,ela} <\sigma_{ela,ine} = \sigma_{ine,ela} <
\sigma_{ine,ine}$.
The two approaches give quite different results.
In the first (naive) approach the inelastic-inelastic contribution is
considerably smaller than the elastic-inelastic or inelastic-elastic ones. 
In the approach when the photon distribution in the proton undergoes QCD 
$\otimes$ QED evolution, it is the inelastic-inelastic contribution 
which is the biggest out of the four contributions.
This shows that including the photon into the evolution equation is crucial.
This is also a lesson for other processes known from
the literature, where photon-photon processes are possible.



\end{document}